\documentclass[conference]{IEEEtran}
\ifCLASSINFOpdf
\else
\fi
\usepackage{graphics}
\usepackage[dvips]{epsfig}
\usepackage{boxedminipage}
\usepackage{rotating}
\usepackage{amssymb,amsmath,amsthm}
\newtheorem{theorem}{Theorem}

\newtheorem{lemma}{Lemma}
\newtheorem{remark}{Remark}
\newtheorem{corollary}{Corollary}

\newcommand{\indp}{\small\raisebox{-0.4mm}{\rotatebox{90}{$\models$}\,}}

\newcommand{\mmse}{\mathrm{mmse}}

\begin{document}
%
\title{Incremental Refinement using a Gaussian\\ Test Channel}

\author{\IEEEauthorblockN{Jan \O stergaard}
\IEEEauthorblockA{Department of Electronic Systems\\
Aalborg University \\
Aalborg, Denmark \\
janoe@ieee.org} \and
\IEEEauthorblockN{Ram Zamir}
\IEEEauthorblockA{
Department  of Electrical Engineering-Systems \\
Tel Aviv University \\
Tel Aviv, Israel \\
zamir@eng.tau.ac.il}
}


%


\maketitle

\begin{abstract}
The additive rate-distortion function (ARDF) was developed in order to universally bound the rate loss in the Wyner-Ziv problem, and has since then been instrumental in e.g., bounding the rate loss in successive refinements, universal quantization, and other multi-terminal source coding settings.
The ARDF is defined as the minimum mutual information over an additive test channel followed by estimation. In the limit of high resolution, the ADRF coincides with the true RDF for many sources and fidelity criterions. In the other extreme, i.e.,  the limit of low resolutions, the behavior of the ARDF has not previously been rigorously addressed.

In this work, we consider the special case of quadratic distortion and
where the noise in the test channel is Gaussian distributed.
We first establish a link to the I-MMSE relation of Guo et
al.~and use this to show that for any source the slope of the ARDF near zero rate, converges to the slope of the Gaussian RDF near zero rate. 
We  then consider the multiplicative rate loss of the ARDF, and show
that for bursty sources it may be unbounded, contrary to the additive
rate loss, which is upper bounded by 1/2 bit for all sources.
We finally show that
unconditional incremental refinement, i.e., where each refinement is encoded independently of the other refinements, is ARDF optimal in the limit of low resolution, independently of the source distribution.  
Our results also reveal under which conditions linear estimation is
ARDF optimal in the low rate regime.
\end{abstract}


%
\IEEEpeerreviewmaketitle

\section{Introduction}

Shannon's rate-distortion function (RDF) for a source $X$ and
distortion measure
$d(\cdot,\cdot)$ is given by 
\begin{equation}\label{eq:rdf}
R(D) = \inf  I(X;Y),
\end{equation}
where the infimum is over all reconstructions $Y$ such that the
expected distortion satisfies $\mathbb{E}[ d(X,Y)] \leq D$. 
Even though~(\ref{eq:rdf}) perhaps appears simple and innocent, it is
well-known that it is generally very hard to explicitly
compute. In fact, there exists only very few cases
where~(\ref{eq:rdf}) is known in closed-form, e.g., Gaussian sources
and MSE, binary sources and Hamming distances etc.
In the information theoretic literature, several methods have been proposed to 
approximate the RDF's e.g., iterative numeric solutions, high-resolution source coding, and (universal) bounds.
In the first case, the Arimoto-Blahut algorithm is able to
numerically obtain the rate-distortion function for arbitrary finite
input/output alphabet sources and single-letter distortion
measures~\cite{blahut:1972}. In the second case, for continuous
alphabet sources, it was shown by Linder and Zamir that the Shannon
lower bound (SLB) is asymptotically tight for norm-based distortion
metrics~\cite{linder:1994}. Thus, at asymptotically high coding rates,
the RDFs can be approximated by simple formulaes. 
In the third case, alternative RDFs, which are easier to compute and analyze, are used to bound the true RDFs. 
For example, at general resolution and for difference distortion measures, the SLB provides a  \emph{lower} bound to the true RDF for many sources. 
On the other hand, Zamir presented in~\cite{zamir:1996} an additive RDF (ARDF), which consists of an additive test channel followed by estimation. 
The ARDF has been shown to be a convenient tool for \emph{upper} bounding the rate loss in many source coding problems. 
In particular, it was shown in~\cite{zamir:1996} that the additive
rate loss in the Wyner-Ziv problem is at most 1/2 bit for all
sources. Similarly, it was shown by Lastras and Berger
in~\cite{lastras:2001}, that the additive rate loss in the successive refinement problem is at most 1/2 bit per stage. The ARDF has also been successfully applied to upper bound the rate loss in other multi-terminal problems, cf.~\cite{zamir:1996,zamir:1999}. In the limit of high resolution, the ARDF coincides with the true RDF for many sources and fidelity criterions~\cite{linder:1994}. In the other extreme, i.e., in the limit of low resolutions, the behavior of the ARDF has not been rigorously addressed. 
There has, however,  been a great interest in the counter part to low resolution source coding, i.e., communication at low SNR, e.g., ultra-wideband communication~\cite{lapidoth:2003}. 
A motivating factor for considering the low SNR regime in
communications, is that the absolute value of the slope of the
capacity-cost function is large (and therefore small for the cost-capacity
function), which indicates that one gets the most channel capacity per
unit cost at low SNR, as was shown by Verd\'u~\cite{verdu:1990}. 
Interestingly, Verd\'u also showed that for rate-distortion at low rates, the most cost effective operating point in terms of bits per unit distortion, is near zero rate~\cite{verdu:1990}. This follows since the absolute value of the slope of the RDF is minimized when the distortion approaches its maximum. 

In this paper, we are interested in analyzing the ARDF at low resolutions. We consider the special case of the ARDF where the test channel's noise is Gaussian and the distortion measure is the MSE. 
We establish a link to the mutual information -- minimum mean squared
estimation (I-MMSE) relation of Guo et al.~\cite{guoshamai:2005} and
use this to show that for any source the slope of the ARDF near zero rate, converges to the slope of the Gaussian RDF near zero rate. 
We then consider the multiplicative rate loss of this ARDF and show that for bursty sources it may be unbounded. 
We also show that unconditional incremental refinement, i.e., where each refinement is encoded independently of the other refinements, 
is ARDF optimal in the limit of low resolution, independently of the source distribution. 
In particular, let an arbitrarily distributed source $X$ be encoded
into $k$ representations $Y_i = \sqrt{\gamma}X + N_i$ where $\{N_i\},i=1,\dotsc, k,$ are mutually independent, Gaussian distributed, and independent of $X$. Then we show that $I(X;Y_1,\dotsc, Y_k)  \approx \sum_i I(X;Y_i)$ at low rates. 
Moreover, the joint reconstruction follows by simple linear estimation of $X$ from $\{Y_1,\dotsc, Y_k\}$. 
If side information $Z$, where $Z$ is independent of $N_i,
i=1,\dotsc,k$, but arbitrarily jointly distributed with $X$, is
available both at the encoder and decoder, we show that
$I(X;Y_1,\dotsc, Y_k|Z)  \approx \sum_i I(X;Y_i|Z)$. In this case,
however, the best conditional estimator $\mathbb{E}[X|Y_1,\dotsc,
Y_k,Z]$ is generally not linear. We provide the exact conditions
for ARDF optimality of linear estimation in the low rate regime.

\section{Background}\label{sec:background}
In this section, we present two existing important concepts that we
will be needing in the sequel, i.e., 
the additive RDF and the I-MMSE
relation.

\subsection{The Additive Rate-Distortion Function}
The additive (noise) RDF, as defined by Zamir
in~\cite{zamir:1996}, describes the best rate-distortion performance achievable
for any additive noise followed by optimum estimation, including
the possibility of time sharing (convexification). 
In the current paper, we restrict attention to Gaussian noise, MMSE
estimation (MSE distortion), and no time-sharing, so we take the ``freedom'' to use
the notation \emph{additive RDF}, $R_X^{\text{add}}(D)$, for this special case (i.e.\ no minimization over free parameters).
Specifically, let $\mathrm{var}(X|Y)$ denote the minimum possible MSE
in estimating $X$ from $Y$, i.e.,
\begin{equation}
\mathrm{var}(X|Y) \triangleq \mathbb{E}[ (\mathbb{E}[X|Y] - X)^2].
\end{equation}
Moreover, let the additive noise $N$ be zero-mean Gaussian
distributed with variance $0<\theta<\infty$. Then,  
\begin{equation}
R_X^{\text{add}}(D) = I(X;X+N),
\end{equation}
where the noise variance $\theta$ is chosen such that $D=\mathrm{var}(X|X+N)$.

\subsection{The I-MMSE Relation}
Using an \emph{incremental} Gaussian channel,
Guo et al.~\cite{guoshamai:2005} was able to establish an explicit
connection between information theory and estimation theory. For
future reference, we include this result below:


\begin{theorem}[\cite{guoshamai:2005}]\label{theo:gsv}
Let $N$ be zero-mean Gaussian of unit variance, independent of $X$, and let $X$ have an arbitrary distribution $P_X$ that satisfies $\mathbb{E}X^2 < \infty$. Then
\begin{equation}\label{eq:immse}
\frac{\mathrm{d}}{\mathrm{d}\gamma} I(X;\sqrt{\gamma} X + N) = \frac{\log_2(e)}{2} \mmse(\gamma),
\end{equation}
where
\begin{equation}
\mmse(\gamma) = \mathbb{E}[ (X - \mathbb{E}[ X| \sqrt{\gamma}X + N]
)^2] = \mathrm{var}(X|\sqrt{\gamma}X+N).
\end{equation}
\end{theorem}

\section{Incremental Refinements}\label{sec:incremental}

\subsection{The Slope of the ARDF}
We will show that the slope of $R_X^{\text{add}}(D)$ at $D=D_{\text{max}}$ for
a source $X$ with variance $\sigma_X^2$ is independent of the
distribution of $X$. In fact, the slope is identical to the slope of the RDF of a Gaussian source $X'$ with
variance $\sigma_{X'}^2=\sigma_X^2$. 
This is interesting since the RDF of any zero-mean source $X$ with a
variance $\mathrm{var}(X) = \sigma_X^2$ meets the Gaussian RDF at
$D=D_{\text{max}}=\sigma_X^2$. Thus, since the Gaussian RDF can be obtained by linear estimation, it follows that $R_X^{\text{add}}(D)$ can also be obtained by linear estimation near $D_\text{max}$. 

\begin{lemma}\label{lem:slope_fx}
Let $Y=\sqrt{\gamma}X + N,$ where $N \indp X$, $X$ is arbitrarily
distributed with variance $\sigma_X^2$ and $N$ is Gaussian
distributed according to $\mathcal{N}(0,1)$. Moreover, let $R_X^{\text{add}}(D)$
be the additive RDF. Then 
\begin{equation}\label{eq:slope}
\lim_{D\to D_{\mathrm{max}}}\frac{\mathrm{d}}{\mathrm{d}D} R_X^{\text{add}}(D)= -\frac{\log_2(e)}{2\sigma_X^2},
\end{equation}
irrespective of the distribution on $X$.
\end{lemma}

\begin{remark}
Interestingly, it was shown by Marco and Neuhoff~\cite{marco:2006} that in the
quadratic memoryless Gaussian case, the operational rate-distortion
function of the scalar uniform quantizer (followed by entropy coding) has the same
slope as~(\ref{eq:slope}). Thus, in this particular case, the optimal scalar
quantizer is as good as any vector quantizer.
\end{remark}

\subsection{Multiplicative Rate Loss in the Low Rate Regime}
Recall that in e.g., the successive refinement problem, the
\emph{additive} rate loss is no more than 0.5 bits per stage. We will
now show
that the \emph{multiplicative} rate loss may be unbounded. 

Let $X$ be a Gaussian mixture source with a density $P_X(x)$ given by
$P_X(x) = P_0 \mathcal{N}(0,\sigma_0^2) + P_1 \mathcal{N}(0,\sigma_1^2)$,
where $P_0+P_1=1$.
The variance $\sigma_X^2$ of $X$ is $\sigma_X^2 = P_0 \sigma_0^2 + P_1\sigma_1^2$.
The components contribution can be parametrized by $\lambda\in [0;1]$ as
follows: $P_0\sigma_0^2 = \lambda \sigma_X^2, P_1\sigma_1^2 = (1-\lambda)\sigma_X^2$.
It will be convenient to let $\sigma_X^2 = 1$ and
$\lambda=\frac{1}{2}$. Moreover, we shall assume that $\sigma_1^2 > 1
> \sigma_0^2 \geq  \frac{1}{2}$.
Notice that as $\sigma_1^2\to\infty$ we have that $P_1\to 0, P_0\to
1$, and $\sigma_0^2\to \frac{1}{2}$.

At this point, let $S = 0$ with probability $P_0$ and $S=1$ with
probability $P_1$, and let $S$ be an indicator of the two components,
i.e., $X \sim \mathcal{N}(0,\sigma_0^2)$, if $S=0$, and
$X\sim \mathcal{N}(0,\sigma_1^2)$,  if $S=1$.
The RDF, conditional on the indicator $S$, is given by
\begin{align*}
R_{X|S}(D) = 
\begin{cases}
\frac{1}{2}\sum_{i\in \{0,1\}} P_i \log_2(\sigma_i^2 / D), &
\text{if $0<D\leq \sigma_0^2$}, \\[5mm]
\displaystyle\frac{P_1}{2}\log_2\bigg( \frac{P_1\sigma_1^2}{D - P_0\sigma_0^2} \bigg), &  \text{if $\sigma_0^2 < D < 1$}.
\end{cases}
\end{align*}
Thus, the slope of $R_{X|S}(D)$ w.r.t.\ $D$ is given by
\begin{align}
\lim_{D\to \sigma_X^2}\frac{\mathrm{d}}{\mathrm{d}D} R_{X|S}(D) &=
 -\frac{P_1}{4\ln(2)\sigma_X^2},
\end{align}
which tends to zero as $\sigma_1^2\to \infty$ and $P_1\to 0$. It
follows from this fact and from Lemma~\ref{lem:slope_fx} that the ratio of the slope of the conditional RDF and the
slope of the ARDF grows unboundedly as $\sigma_1^2\to
\infty$. Moreover, as $\sigma_1^2\to \infty$, $\sigma_0^2\to
\frac{1}{2}$, which implies that it becomes increasingly easier for the uninformed encoder/decoder to
guess the correct component of the source. Thus, the conditional RDF
converges towards the true RDF $R_X(D)$, from which it follows that the
ratio $\lim_{\sigma_1^2/\sigma_0^2\to \infty}\lim_{D\to \sigma_X^2} R_X^{\text{add}}(D)/R_X(D)\to\infty$.\footnote{A rigorous proof of the
  convergence is omitted due to space considerations.} 

\subsection{Unconditional Incremental Refinements}
We will now show that unconditional incremental refinement, i.e., where each refinement is encoded independently of the other refinements, 
is ARDF optimal in the limit of low resolution, independently of the
source distribution. This result is not only of theoretical value but
is also useful in practice, since conditional source coding is
generally more complicated than unconditional source coding, i.e., 
creating descriptions that are individually optimal and at the same time jointly
optimal is a long standing problem in information theory, where it is known as the
multiple descriptions problem~\cite{elgamal:1982}.

\begin{lemma}\label{lem:oversampling}
Let $X$ be arbitrarily distributed with
variance $\sigma_X^2$, and let $N_i\indp X, i=0,\dotsc, k-1,$ be a sequence
of zero-mean mutually independent Gaussian sources each with variance $\sigma_N^2$. Then
\begin{equation*}
I(X;X+N_0, \dotsc, X+N_{k-1}) = I(X;X+ \frac{1}{\sqrt{k}} N_0).
\end{equation*}
\end{lemma}

\begin{lemma}\label{lem:uncond}
Let $Y_i = \sqrt{\gamma}X + N_i, i =0,\dotsc, k-1$, where $N_i\indp X,
\forall i$. Moreover, let $X$ be arbitrarily
distributed with variance $\sigma_X^2$ and let $N_0,\dotsc, N_{k-1},$
be zero-mean unit-variance i.i.d.\ Gaussian distributed.  Then
\begin{align*}
\lim_{\gamma \to 0}\frac{1}{\gamma} I(X;Y_0,\dotsc, Y_{k-1}) &= k
\lim_{\gamma \to 0}\frac{1}{\gamma} I(X;\sqrt{\gamma}X + N_0) \\
&= \frac{k\log_2(e)}{2}\sigma_X^2
\end{align*}
and
\begin{equation}\label{eq:jointd}
\lim_{\gamma \to 0}\frac{1}{\gamma}\bigg[\frac{1}{\mathrm{var}(X|Y_1,\dotsc,Y_{k-1})} -\frac{1}{\sigma_X^2}\bigg] = k.
\end{equation}
\end{lemma}

\begin{figure}
\begin{center}
\includegraphics[width=8cm]{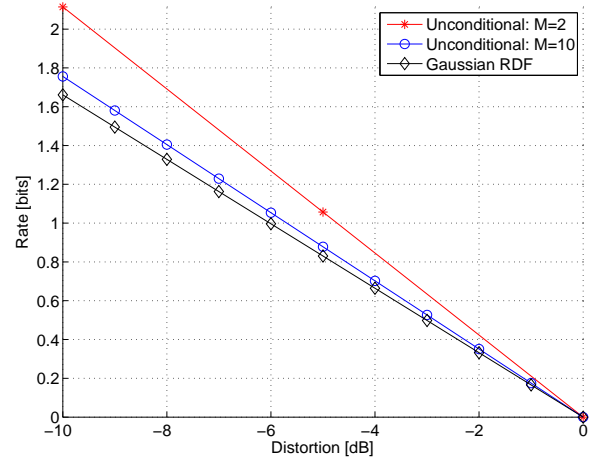}
\vspace{-2mm}
\caption{Unconditional and conditional successive refinements in the
  quadratic Gaussian case.}
\label{fig:uncond1}
\end{center}
\vspace{-8mm}
\end{figure}

To illustrate the importance of Lemma~\ref{lem:uncond}, let us consider the situation of a zero-mean unit-variance memoryless
Gaussian source $X$, which is to be encoded successively in $M$
stages. In stage $i$, $L$ descriptions $Y_{i,j}, j=1,\dotsc, L$, are
constructed unconditionally of each other. Thus, for the same coding
rate (at each stage), the joint distortion
$\mathrm{var}(X|Y_{1,1},\dotsc,Y_{1,L},
\dotsc,Y_{i,1},\dotsc,Y_{i,L})$
in the $i$th stage is worse than if only a single joint description
within each stage had been created. In 
fact, in the symmetric case where all individual descriptions within
stage $i$ has the same distortion $d_i$ and rate $r_i$, 
it can be shown that the joint distortion $D_i$ of the $i$th stage is
given by
\begin{equation}\label{eq:Dsuc}
D_i  = \frac{d_i}{L - (L-1)d_i / D_{i-1}}
\end{equation}
and the sum-rate at stage $i$ is given by
\begin{equation}\label{eq:Rsuc}
R_i = L\sum_{j=1}^i  r_i = L  \sum_{j=1}^i  \frac{1}{2} \log_2( D_{j-1} / d_j ),
\end{equation}
where $D_0=d_0=\sigma_X^2$.
Since the Gaussian
source is successively refinable, using conditional refinements will
achieve the true RDF given by $R_i^* = \frac{1}{2}\log_2( 1/D_i )$,
where $D_i$ is given by~(\ref{eq:Dsuc}). On the other hand, the rate
required when unconditional coding is used is given
by~(\ref{eq:Rsuc}). For comparison, we have illustrated the
performance of unconditional and conditional coding when the source is
encoded into $L=2$ descriptions per stage, 
for the case of 
of $M=2$ and $M=10$ increments (stages), respectively, see
Fig.~\ref{fig:uncond1}. In this example,
$\sigma_X^2=1$ and $D_M = 0.1$. Notice that when using smaller
increments, i.e., when $M=10$ as compared to when $M=2$, the resulting
rate loss due to using
unconditional coding is significantly reduced. 

\subsection{Unconditional Incremental Refinements (Side Information)}\label{sec:incremental_side}
The case of additional side information available at the encoder and
the decoder was not considered by Guo et
al.~in~\cite{guoshamai:2005}. Below we generalize
Theorem~\ref{theo:gsv} to include side information:
\begin{lemma}\label{lem:condmi}
Let $Y=\sqrt{\gamma}X+N$ where $N\sim \mathcal{N}(0,1)$ and $X$ is
arbitrarily distributed, independent of $N$ and of variance
$\sigma_X^2$. 
Let $Z$ be arbitrarily
distributed and correlated with $X$ but independent of $N$. Then
\begin{equation}
\lim_{\gamma\to 0} \frac{1}{\gamma}I(X;Y|Z) = \frac{\log_2(e)}{2}\mathrm{var}(X|Z).
\end{equation}
\end{lemma}

\begin{corollary}
Let $Y_i = \sqrt{\gamma}X + N_i, i =0,\dotsc, k-1$, where $N_i\indp X,
\forall i,$ and $N_0,\dotsc, N_{k-1}$. Let $X$ be arbitrarily
distributed with variance $\sigma_X^2$ and let $N_0,\dotsc, N_{k-1},$
be zero-mean unit-variance i.i.d.\ Gaussian distributed. 
Let $Z$ be arbitrarily
distributed and correlated with $X$ but independent of $N_i,\forall i$. Then
\begin{align*}
\lim_{\gamma \to 0}\frac{1}{\gamma} I(X;Y_0,\dotsc, Y_{k-1}|Z) &= \frac{k\log_2(e)}{2}\mathrm{var}(X|Z).
\end{align*}
\end{corollary}

\subsection{Conditions for Optimality of Linear Estimation}
It was recently shown by Akyol et al.~\cite{akyol:2010}, that for an
arbitrarily distributed source $X$, contaminated by Gaussian noise $N$, the
MMSE estimator of $X$ given $Y=\sqrt{\gamma}X+N$, converges in
probability to a linear estimator, in the limit where $\gamma\to 0$. 
Contrary to this result, we show that the conditional MMSE
estimator 
$\mathbb{E}[X|Y,Z]$ with side information $Z$, 
where $Z$ is independent of $N$ but is arbitrarily correlated with
$X$ is generally not linear.

\begin{lemma}\label{lem:nonlinear}
Let $Y=\sqrt{\gamma}X + N,$ where $N \indp X$, $X$ is arbitrarily
distributed with variance $\sigma_X^2$ and $N$ is Gaussian
distributed according to $\mathcal{N}(0,1)$. Moreover, let $Z$ be
arbitrarily distributed, independent of $N$ but arbitrarily correlated
with $X$.  Then the conditional MMSE estimator $\mathbb{E}[X|Y,Z]$ is
linear if and only if 
\begin{equation}\label{eq:mmse_equiv}
\mathbb{E}[\mathrm{var}(X|Z=z)^2] = \mathrm{var}(X|Z)^2,
\end{equation}
where 
\begin{equation*}
\mathbb{E}[\mathrm{var}(X|Z=z)^2] \triangleq 
\mathbb{E}_Z[ (\mathbb{E}_X[ (\mathbb{E}_X[X|Z=z] - X)^2] )^2]
\end{equation*}
and
\begin{equation*}
\mathrm{var}(X|Z)^2 \triangleq (\mathbb{E}_X[ (\mathbb{E}_X[X|Z] - X)^2])^2.
\end{equation*}
\end{lemma}

In the case where $X, Z$ are jointly Gaussian, it is easy to show
that~(\ref{eq:mmse_equiv}) is satisfied and, thus, the MMSE estimator $\mathbb{E}[X|Y,Z]$
is trivially linear in both $Z$ and $N$. 

\section*{Acknowledgment}
The authors would like to thank Uri Erez who initially proposed the
idea of incremental refinements in the context of multiple
descriptions with feedback.

\appendix

\begin{IEEEproof}[Proof of Lemma~\ref{lem:slope_fx}]
The additive RDF is defined parametrically as $R_X^{\text{add}}(D)$, by
$R_X^{\text{add}}(\gamma) = I(\gamma), 
D(\gamma) = \mmse(\gamma)$,
which implies that
\begin{equation}
R_X^{\text{add}}(D(\gamma))= I(D(\gamma)).
\end{equation}
From the derivative of a composite function, it follows that 
\begin{equation}
\frac{\mathrm{d}}{\mathrm{d}D} R_X^{\text{add}} =
\frac{\frac{\mathrm{d}}{\mathrm{d}\gamma} R_X^{\text{add}}}{
\frac{\mathrm{d}}{\mathrm{d}\gamma} D
}.
\end{equation}
We know that $I(\gamma) =I(X;\sqrt{\gamma}X + N)$ can be expanded as~\cite{guoshamai:2005}
\begin{equation}\label{eq:ig}
\begin{split}
&I(\gamma) = \log_2(e)\bigg[\frac{1}{2}\gamma\sigma_X^2 -\frac{1}{4}\gamma^2\sigma_X^4 +
\frac{1}{6}\gamma^3\sigma_X^6 \\
&- \frac{1}{48}\bigg[ (\mathbb{E}X^4)^2 -
  6\mathbb{E}X^4 - 2(\mathbb{E}X^3)^2 + 15 \bigg]\gamma^4\sigma_X^8 + \mathcal{O}(\gamma^5)\bigg],
\end{split}
\end{equation}
and that
\begin{equation}\label{eq:mmse_g}
\mmse(\gamma) = \sigma_X^2-\gamma\sigma_X^4 +
\gamma^2\sigma_X^4 +\frac{1}{6}\gamma^3\sigma_X^6 + \mathcal{O}(\gamma^4).
\end{equation}
It follows from~(\ref{eq:ig}) that 
\begin{equation}
\lim_{\gamma\to 0}
\frac{\mathrm{d}^2}{\mathrm{d}\gamma^2}  I(\gamma) = -\frac{\log_2(e)}{2}\sigma_X^4.
\end{equation}
From~\cite{guoshamai:2005}, $\lim_{\gamma\to 0}\frac{\mathrm{d}}{\mathrm{d}\gamma}
D = 2\frac{\mathrm{d}^2}{\mathrm{d}\gamma^2}  I(\gamma) =
-\sigma_X^4$. Moreover, since $\lim_{\gamma\to 0}\frac{\mathrm{d}}{\mathrm{d}\gamma}
R_X^{\text{add}} = \frac{\log_2(e)}{2}\mmse(\gamma)=\log_2(e)\sigma_X^2/2$ and since $\gamma\to 0$ implies $D\to D_{\mathrm{max}}$, we have that the slope of $R_X^{\text{add}}(D)$ with respect to
$D$ at $D=D_\mathrm{max}$ is 
\begin{equation}
\lim_{D\to D_{\mathrm{max}}}\frac{\mathrm{d}}{\mathrm{d}D}R_X^{\text{add}} = -\frac{\log_2(e)}{2\sigma_X^2}.
\end{equation}

\end{IEEEproof}

\begin{IEEEproof}[Proof of Lemma~\ref{lem:oversampling}]
Let $Y_i = X+N_i, i=0,\dotsc,k-1,\bar{Y} = [Y_0, \dotsc, Y_{k-1}]^T,$ and let $\bar{Z}$ be the DFT of $\bar{Y}$, i.e.,
\begin{equation}
\bar{Z}_j = \frac{1}{k}\sum_{i=0}^{k-1} Y_i \exp(2\pi ij/k), \quad j = 0,\dotsc, k-1.
\end{equation}
The DC term is given by $Z_0 = X + \frac{1}{k}\sum_{i=0}^{k-1} N_i$. 
The other terms, i.e., $Z_j, j>0$, are AC terms and do not contain $X$ (since $X$ is DC). The AC terms are orthogonal to the DC component of the noise, i.e., $(Z_0-X) \perp Z_j, j>0$, and since the Gaussianity of the noise implies independence, we are left with only the DC term.  
Since the $N_i$'s are mutually independent, the resulting sum-noise component $\frac{1}{k}\sum_{i=0}^{k-1}N_i$ of the DC term has variance $\sigma_N^2/k$. Thus, the DC term is equivalent to $X + \frac{1}{\sqrt{k}}N$, where $N$ is distributed as $N_i$. This shows that $I(X; \bar{Y}) = I(X; Z_0) = I(X; X + \frac{1}{\sqrt{k}} N_0)$. The lemma is proved.
\end{IEEEproof}

\begin{IEEEproof}[Proof of Lemma~\ref{lem:uncond}]
From Lemma~\ref{lem:oversampling}, it is clear that 
\begin{equation}
I(X;Y_0, \dotsc, Y_{k-1}) = I(X;X+ \frac{1}{\sqrt{k}} N_0).
\end{equation}
To get to the standard form with unit-variance noise, we may scale
both $X$ and $\sqrt{\gamma}X + \frac{1}{\sqrt{k}} N_0$ by $\sqrt{k}$
without affecting their mutual information, i.e.,
\begin{equation}
 I(X;X+ \frac{1}{\sqrt{k}} N_0) =  I(\sqrt{k}X;\sqrt{k} X + N_0).
\end{equation}
At this point we use that~\cite{guoshamai:2005}
\begin{equation}
\lim_{\gamma\to 0}\frac{1}{\gamma}I(X';\sqrt{\gamma}X' + N) = \frac{\log_2(e)\sigma_{X'}^2}{2},
\end{equation}
where $X' = \sqrt{k}X$ and $\sigma_{X'}^2 = k\sigma_X^2$. This proves
the first part of the lemma. 
By using well-known linear estimation theory, it is
easy to show that
\begin{align}\notag
\frac{1}{\mathrm{lmmse}(X|Y_1,\dotsc,Y_{k-1})} &=
\frac{1}{\mathrm{var}( X)} + \frac{\gamma}{\mathrm{var}(\frac{1}{\sqrt{k}}N_0)}
\\ \label{eq:jointd1}
&= \frac{1}{\sigma_X^2} + \gamma k,
\end{align}
where $\mathrm{lmmse}{(X|Y_1,\dotsc,Y_{k-1})}$ denotes the MSE due to
estimating $X$ from $Y_1,\dotsc,Y_{k-1}$ using linear estimation.
We now invoke the fact that linear estimation is optimal in the limit
$\gamma\to 0$ and re-order the terms in~(\ref{eq:jointd1}) to get~(\ref{eq:jointd}).
\end{IEEEproof}

\begin{IEEEproof}[Proof of Lemma~\ref{lem:condmi}]
We will extend the proof technique used in~\cite[Lemma
1]{guoshamai:2005} to allow for arbitrary conditional
distributions. To do this, we make use of the fact $Y-X -Z$ forms a Markov
chain (in that order), which will allow us to simplify the
decomposition of their joint distribution.

Let
$\mathbb{E}_\Xi$ denote expectation with respect to $\Xi$. We first
expand the conditional mutual information in terms of the Divergence, i.e.
\begin{align}
I(X;Y|Z) & = \mathbb{E}_Z D(P_{XY|Z} || P_{X|Z} P_{Y|Z}) \notag \\
&= \mathbb{E}_{Z,\{X|Z\}} D(P_{Y|Z,X} || P_{Y|Z})  \notag\\ 
&= \mathbb{E}_{Z,\{X|Z\}} \big[ D(P_{Y|X} || P_{Y'|Z'}) \notag \\ \label{eq:divsplit}
&\quad - D(P_{Y|Z} || P_{Y'|Z'}) \big],
\end{align}
where $P_{Y'|Z'}$ can be chosen arbitrary as long as $D(P_{Y|X} ||
P_{Y'|Z'})$ and $D(P_{Y|Z} || P_{Y'|Z'})$ are both well-defined. Let
$Y'|Z' \sim \mathcal{N}(\sqrt{\gamma}\,\mathbb{E}[X|Z], 1+ \gamma\, \mathrm{var}(X|Z))$.

The first term in~(\ref{eq:divsplit}) is the Divergence between two Gaussian distributions, since $\mathbb{E}[Y|Z,X] = \mathbb{E}[Y|X] = N$ is Gaussian distributed and $\mathbb{E}[Y'|Z']$ is Gaussian since a linear combination of Gaussians remain Gaussian.
In this case we have~\cite{guoshamai:2005}
\begin{align}\notag
\lim_{\gamma\to 0}\frac{1}{\gamma}D(\mathcal{N}(0,\sigma_1^2) ||
\mathcal{N}(0,\sigma_0^2)) &= \lim_{\gamma\to 0}\frac{1}{\gamma}
\log(1+\gamma\, \mathrm{var}(X|Z)) \\
&= \mathrm{var}(X|Z),
\end{align}
where we used that $\lim_{\gamma\to 0} \frac{1}{\gamma}\log(1+\gamma c) = c$.

We now look at the second expression in (\ref{eq:divsplit}) and
use the Markov condition to get to
$P_{Y|Z} = \mathbb{E}_{X|Z}[ P_{Y|X,Z}] = \mathbb{E}_{X|Z}[ P_{Y|X}]$.
With this, we may adapt the proof technique of~\cite{guoshamai:2005}
to obtain:
{\allowdisplaybreaks
\begin{align*}
&\log\left( \frac{ P_{Y|Z}(y|z) } {P_{Y'|Z'}(y|z)} \right) \\
&=
\log\bigg(
\frac{
\frac{1}{\sqrt{2\pi \sigma_N^2}}
\mathbb{E}_{X|Z=z}\bigg[
\exp\big(-\frac{1}{2\sigma_N^2}(y 
-\sqrt{\gamma}X)^2\big)\bigg]
}
{
\frac{1}{\sqrt{2\pi \sigma_0^2}}
 \exp\big(-\frac{1}{2\sigma_0^2}(y
  - \gamma \mathbb{E}[X|z])^2\big)
}
\bigg) \\
 &= 
 \log\bigg(
\mathbb{E}_{X|Z=z}\bigg[\exp\bigg\{ \frac{1}{2\sigma_0^2}(y 
- \mathbb{E}[X|z])^2 \\
&\qquad
 -\frac{1}{2\sigma_N^2}(y-\sqrt{\gamma}X)^2  \bigg\}\bigg]
 \bigg)  + \frac{1}{2}\log\bigg(\frac{\sigma_0^2}{\sigma_N^2}\bigg) \\
&=
 \log\bigg(
\mathbb{E}_{X|Z=z}\bigg[\exp\bigg\{ 
\frac{(y - \sqrt{\gamma}\,\mathbb{E}[X|z])^2}{2(1+ \gamma\,\mathrm{var}(X|z))} \\
&\qquad
 -\frac{(y-\sqrt{\gamma}X)^2}{2\sigma_N^2}  \bigg\}\bigg]
 \bigg)  + \frac{1}{2}\log\bigg(\frac{\sigma_0^2}{\sigma_N^2}\bigg) \\
&\overset{(a)}{=}
\log\bigg(
\mathbb{E}_{X|Z=z}\bigg[
1 +\sqrt{\gamma} y(X-\mathbb{E}[X|z]) \\
&\quad
+\frac{\gamma}{2}(y^2(X-\mathbb{E}[X|z])^2 - y^2\mathrm{var}(X|z) \\
&\quad 
-X^2 + \mathbb{E}[X|z]^2  + o(\gamma)\bigg] 
\bigg) 
+\frac{1}{2}\log(1+\gamma\, \mathrm{var}(X|z)
) \\
&= 
\log(1-\frac{\gamma}2\, \mathrm{var}(X|z) )
+\frac{1}{2}\log(1+\gamma\, \mathrm{var}(X|z)) + o(\gamma) \\
&= o(\gamma),
\end{align*}
where $(a)$ follows by using a series expansion of $\exp(\cdot)$ in
terms of $\gamma$.}
We have thus established that the second term of~(\ref{eq:divsplit})
goes to zero (as a function of $\gamma$) faster than the first
term. Thus, the first term dominates the conditional mutual
information for small $\gamma$. This completes the proof.
\end{IEEEproof}

\begin{IEEEproof}[Proof of Lemma~\ref{lem:nonlinear}]
We first consider the unconditional case, where $Z=\emptyset$. Let us
assume that $\mathbb{E}X = \mu_X \neq 0$. Recall that $Y =
\sqrt{\gamma}X + N$, where $\mathbb{E}N=0$ and $\sigma_N^2=1$. 
For small $\gamma$, the optimal estimator is linear, and we have that
\begin{equation}\label{eq:linest}
\mathbb{E}[X|Y] \approx \mu_X + \alpha(Y - \mu_X),
\end{equation}
where $\alpha$ is the Wiener coefficient given by 
$\alpha = \mathbb{E}[XY]=\sqrt{\gamma}\sigma_X^2$.
From~(\ref{eq:mmse_g}), we know that the MMSE behaves as:
\begin{equation}\label{eq:var_uncond}
\mathrm{var}(X|Y) \approx \sigma_X^2 - \gamma \sigma_X^4.
\end{equation}
On the other hand, in the conditional case with side information $Y$,
for each $Z=z$ the source has mean $\mathbb{E}[X|Z=z]$ and variance
$\mathrm{var}(X|Z=z)$. Using this in~(\ref{eq:linest}), and fixing $Z=z$, leads to
\begin{equation}
\mathbb{E}[X|Y,Z=z] \approx \mathbb{E}[X|Z=z] + \alpha_z (Y - \mathbb{E}[X|Z=z]),
\end{equation}
where the Wiener coefficient depends on  $z$, i.e.,
$\alpha_z=\sqrt{\gamma}\mathrm{var}(X|z)$. 
Using~(\ref{eq:mmse_g}) for a fixed $Z=z$ yields
\begin{equation}
\mathrm{var}(X|Y,Z=z) \approx \mathrm{var}(X|Z=z) - \gamma\, \mathrm{var}(X|Z=z)^2.
\end{equation}
Taking the average over $Z$ results in
\begin{equation}\label{eq:var_cond}
\mathrm{var}(X|Y,Z) \approx \mathrm{var}(X|Z) - \gamma\,\mathbb{E}_Z[\mathrm{var}(X|Z=z)^2],
\end{equation}
where $\mathrm{var}(X|Z)\triangleq \mathbb{E}_Z[\mathrm{var}(X|Z=z)]$. By Jensen's
inequality, it follows that
\begin{equation}
\mathbb{E}_Z[ \mathrm{var}(X|Z=z) ^2] \geq \mathrm{var}(X|Z)^2,
\end{equation}
with equality if and only if the conditional variance $\mathrm{var}(X|Z=z)$ is
independent of the realization of $z$.
Thus, comparing~(\ref{eq:var_cond})
to~(\ref{eq:var_uncond}) shows that the linear estimator is generally
not optimal.

\end{IEEEproof}

\vspace{-5mm}

\bibliographystyle{IEEEtran}





%

\end{document}